\documentclass[prl,a4paper,twocolumn,showpacs,floatfix,superscriptaddress,amsmath,amsfonts,amssymb,preprintnumbers,citeautoscript]{revtex4}
\pdfoutput=1
\usepackage[T1]{fontenc}\usepackage[latin1]{inputenc}
\usepackage{dcolumn,graphicx,color,booktabs,microtype,afterpage} \graphicspath{{Figures/}}
\usepackage[charter,greekuppercase=italicized]{mathdesign}\usepackage{sidecap}
\renewcommand{\figurename}{Fig.}
\renewcommand{\tablename}{Table}
\makeatletter\renewcommand{\fnum@figure}[1]{\figurename~\thefigure~(color online).}\makeatother
\makeatletter\renewcommand{\fnum@table}[1]{\tablename~\thetable.}\makeatother

\newcount\hh \newcount\mm
\hh=\time \divide\hh by 60
\mm=\hh \multiply\mm by 60 \mm=-\mm
\advance\mm by \time
\def\now{\number\hh:\ifnum\mm<10{}0\fi\number\mm}

\usepackage[colorlinks,plainpages=false,linkcolor=blue,urlcolor=blue,citecolor=blue,pdfpagemode=UseNone,pdfstartview=FitBH]{hyperref}

\begin{document}

\makeatletter\renewcommand{\ps@plain}{%
\def\@evenhead{\hfill\itshape\rightmark}%
\def\@oddhead{\itshape\leftmark\hfill}%
\renewcommand{\@evenfoot}{\hfill\small{--~\thepage~--}\hfill}%
\renewcommand{\@oddfoot}{\hfill\small{--~\thepage~--}\hfill}%
}\makeatother\pagestyle{plain}


\title{Incommensurate short-range multipolar order parameter of phase~II in Ce$_3$Pd$_{20}$Si$_6$}

\author{P.~Y.~Portnichenko}
\affiliation{Institut für Festkörperphysik, TU Dresden, D-01069 Dresden, Germany}

\author{S.~Paschen}
\affiliation{Institute of Solid State Physics, Vienna University of Technology, Wiedner Hauptstr. 8--10, A-1040 Vienna, Austria}

\author{A.~Prokofiev}
\affiliation{Institute of Solid State Physics, Vienna University of Technology, Wiedner Hauptstr. 8--10, A-1040 Vienna, Austria}

\author{M.~Vojta}
\affiliation{Institut für Theoretische Physik, TU Dresden, D-01062 Dresden, Germany}

\author{A.~S.~Cameron}
\affiliation{Institut für Festkörperphysik, TU Dresden, D-01069 Dresden, Germany}

\author{J.-M. Mignot}
\affiliation{Laboratoire L\'{e}on Brillouin, CEA-CNRS, CEA/Saclay, F-91191 Gif sur Yvette, France}

\author{A.\,Ivanov}
\affiliation{Institut Laue-Langevin, 6 rue Jules Horowitz, F-38042 Grenoble Cedex 9, France}

\author{D.~S.~Inosov}\email[Corresponding author: \vspace{5pt}]{Dmytro.Inosov@tu-dresden.de}
\affiliation{Institut für Festkörperphysik, TU Dresden, D-01069 Dresden, Germany}

\begin{abstract}\parfillskip=0pt\relax
\noindent The clathrate compound Ce$_3$Pd$_{20}$Si$_6$ is a heavy-fermion metal that exhibits magnetically hidden order at low temperatures. Reputedly, this exotic type of magnetic ground state, known as ``phase~II'', could be associated with the ordering of Ce\,4$f$ quadrupolar moments. In contrast to conventional (dipolar) order, it has vanishing Bragg intensity in zero magnetic field and, as a result, has escaped direct observation by neutron scattering until now. Here we report the observation of diffuse magnetic neutron scattering induced by an application of magnetic field along either the $[1\overline{1}0]$ or the $[001]$ direction within phase~II. The broad elastic magnetic signal that surrounds the (111) structural Bragg peak can be attributed to a short-range G-type antiferromagnetic arrangement of field-induced dipoles modulated by the underlying multipolar order on the simple-cubic sublattice of Ce ions occupying the 8$c$ Wyckoff site. In addition, for magnetic fields applied along the $[001]$ direction, the diffuse magnetic peaks in Ce$_3$Pd$_{20}$Si$_6$ become incommensurate, suggesting a more complex modulated structure of the underlying multipolar order that can be continuously tuned by a magnetic field.
\end{abstract}

\keywords{heavy-fermion compounds, multipolar ordering, incommensurate order, elastic neutron scattering}
\pacs{75.20.Hr 71.27.+a 75.25.-j 61.05.fg\vspace{-3pt}}

\maketitle

\section{I. Introduction}\vspace{-5pt}

Some heavy-fermion materials show so-called hidden-order phases, which are invisible to conventional diffraction techniques, and whose microscopic origin remained controversial for
decades. Such hidden-order phases have been observed in a variety of compounds containing $4f$ and $5f$ elements, for example, URu$_2$Si$_2$ \cite{MydoshOppeneer11, MydoshOppeneer14}, NpO$_2$ \cite{PaixaoDetlefs02}, skutterudites~\cite{YogiNiki09,AokiSanada08}, YbRu$_2$Ge$_2$ \cite{JeevanGeibel06}, or CeB$_6$ \cite{EffantinRossatMignod85, RossatMignod87, CameronFriemel16}. It is often assumed that the multipolar moments of the $f\!$ electrons in their specific crystal-field environment play a decisive role in the formation of these phases \cite{ThalmeierTakimoto08, KuramotoKusunose09, SantiniCarretta09}. The competition or coexistence of multipolar ordering (MPO) with more conventional magnetic order parameters, such as ferro- or antiferromagnetism, gives rise to complex magnetic-field\,--\,temperature phase diagrams in these compounds, often with multiple quantum critical points \cite{WengSmidman16, PaschenLarrea14}, that provide a rich playground for experimental and theoretical investigations.

\begin{figure}[b!]\vspace{-0.3em}
\begin{center}
\includegraphics[width=0.86\columnwidth]{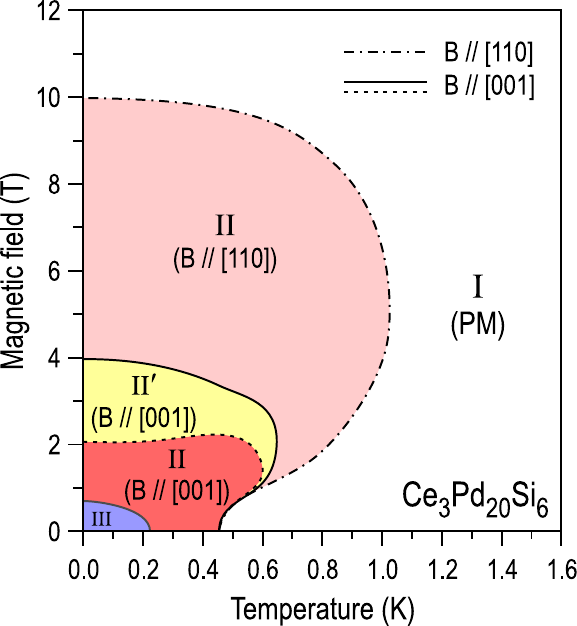}\vspace{-0.5em}
\end{center}
\caption{Schematic phase diagram of Ce$_3$Pd$_{20}$Si$_6$ after Refs.\,\citenum{GotoWatanabe09, MitamuraTayama10, OnoNakano13}, showing strong anisotropy of the hidden-order phase~II with respect to the magnetic field direction. Phase I represents the paramagnetic state, phase~II is associated with antiferroquadrupolar order, phase~III is the incommensurate dipolar magnetic phase~\cite{CustersLorenzer12, LorenzerPhD12, DeenUnpublished}, whereas the order parameter of phase~II$^\prime$ that appears only for $\mathbf{B}\parallel[001]$ still remains unclear.\vspace{-3pt}}
\label{Fig:PhaseDiagram}
\end{figure}

Because the multipolar ordering is often coupled to another magnetically ordered state, the instances where the orbital ordering appears as a separate solitary phase are rare. Several Ce-based compounds, such as CeB$_6$, Ce$_{3}$Pd$_{20}$Si$_{6}$, and Ce$_{3}$Pd$_{20}$Ge$_{6}$, are of particular interest as they display such standalone MPO phases \cite{PaschenLarrea14, KitagawaTakeda96, NemotoYamaguchi03}. Close inspection of the magnetic phase diagram for Ce$_{3}$Pd$_{20}$Si$_{6}$ \cite{GotoWatanabe09, MitamuraTayama10, OnoNakano13}, as schematically shown in Fig.~\ref{Fig:PhaseDiagram}, suggests that it shows remarkably similar behavior to that of CeB$_{6}$, but with reduced temperature and field scales~\cite{PaschenLarrea14, ProkofievCusters09}. Structurally, the $R_{3}\mathrm{Pd}_{20} X_{6}$ ($R$\,=\,rare earth, $X$\,=\,Si, Ge) compounds are far more complex than CeB$_{6}$, because they host two interpenetrating Ce sublattices. One of them, corresponding to the 8$c$ \mbox{Wyckoff} position, possesses a simple-cubic structure like in CeB$_6$, whereas the other one, formed by atoms on the 4$a$ Wyckoff position, has a geometrically frustrated face-centered-cubic (fcc) structure, with each Ce ion being surrounded by either Pd/Ge or Si-Pd/Si-Ge nonmagnetic ``cages''. This unprecedented coexistence in the same material of two inequivalent Kondo lattices with different symmetry has been considered theoretically \cite{BenlagraFritz11}, and the two sublattices were predicted to exhibit two drastically different Kondo temperatures due to the competitive Kondo-screening effects. This is consistent with our recent neutron-scattering observation suggesting that quasielastic magnetic scattering in Ce$_{3}$Pd$_{20}$Si$_{6}$ originates from magnetic correlations on the simple-cubic Ce\,8$c$ site only, while the 4$a$ site remains ``magnetically silent''~\cite{PortnichenkoCameron15}. This could be due to either frustration on the fcc sublattice, strong Kondo screening of their magnetic moments, or perhaps both effects combined.

\begin{figure*}[t]\vspace{-6pt}
\includegraphics[width=\textwidth]{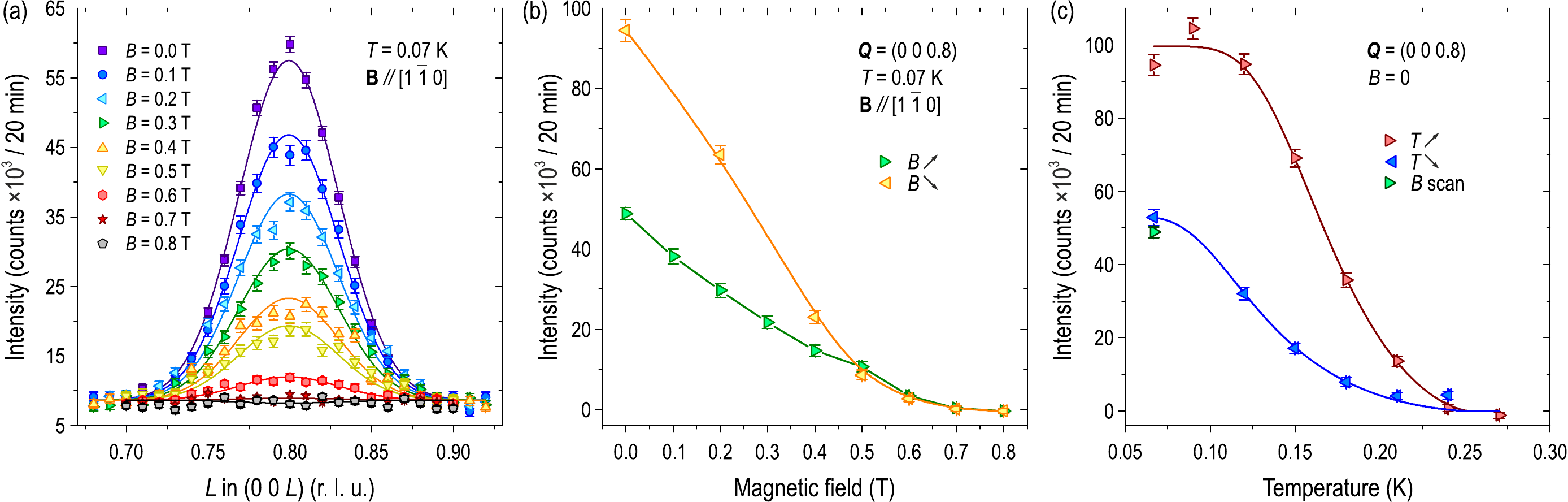}\vspace{-3pt}
\caption{Magnetic-field and temperature dependence of the incommensurate $(0~0~0.8)$ magnetic Bragg peak that represents the order parameter of phase~III. (a)~Unprocessed elastic neutron scattering data measured at $T=0.07$\,K for various fields applied along $[1\overline{1}0]$, fitted to a Gaussian model to extract the peak intensity. (b)~Field dependence of the background-subtracted magnetic Bragg peak amplitude for increasing~($\triangleright$) and decreasing~($\triangleleft$) field, $\mathbf{B}\parallel[1\overline{1}0]$. The observed hysteresis below 0.5~T is a result of domain selection. (c)~Temperature dependence of the same amplitude for increasing~($\triangleright$) and decreasing~($\triangleleft$) temperature that was measured in zero field after the application of a 0.8~T field. The point labeled ``$B$ scan'' is the first point of the ``$\kern.7pt\triangleright$'' dataset in panel (b), measured before the field was applied. All data points in panels (b) and (c) were obtained from Gaussian fits similar to those shown in panel~(a).}
\label{Fig:AFM_110}
\end{figure*}

Up to now, the true order parameter and the propagation vector of phase~II in Ce$_{3}$Pd$_{20}$Si$_{6}$ have remained elusive and have not been observed directly by any diffraction method. Magnetization and specific-heat measurements revealed strong anisotropy of this phase with respect to the direction of applied magnetic field: For $\mathbf{B}\parallel[001]$ it is suppressed by as little as 2~T, for $\mathbf{B}\parallel[110]$ this critical field, $B_{\rm II}$, dramatically increases to 10~T, whereas for $\mathbf{B}\parallel[111]$ this phase persists to even higher fields ($\sim$\,16\,--\,18~T by extrapolation) \cite{MitamuraTayama10, OnoNakano13}. Moreover, only for $\mathbf{B}\parallel[001]$ the system exhibits another transition to the phase~II$^\prime$ that is stabilized between 2 and 4~T (see Fig.\,\ref{Fig:PhaseDiagram}), whose microscopic origin remains unknown.

\vspace{-5pt}\section{II. Experimental Results}\vspace{-5pt}

Here we employ elastic neutron scattering to reveal the order parameter of phase~II in Ce$_3$Pd$_{20}$Si$_6$ for the first time. All measurements were taken using the cold-neutron triple-axis spectrometers \textsc{4F2} (Laboratoire L\'eon Brillouin, Saclay, France) and \textsc{Thales} (Institut Laue-Langevin, Grenoble, France). The energy analysis was used to separate the weak elastic-scattering signal from inelastic contributions. Both spectrometers were operated with the fixed final neutron wave vector $k_\text{f}$ set to 1.3\,\AA$^{-1}$ and a cold beryllium filter installed between the sample and the analyzer to suppress higher-order contamination from the monochromator. In all the reported experiments, we used the same sample as in Ref.\,\citenum{PortnichenkoCameron15}, consisting of two coaligned single crystals with a total mass of $\sim$\,5.9\,g and a mosaic spread better than 0.5$^\circ\!$. The crystals were mounted on a copper sample holder in the $(HHL)$ scattering plane in a $^3$He/$^4$He dilution refrigerator inside a cryomagnet. All measurements with magnetic field along the $[1\overline{1}0]$ direction were done at \textsc{4F2} using a vertical-field 9~T magnet, whereas the measurements with magnetic field along $[001]$ were performed at \textsc{Thales} using a horizontal-field 3.8~T magnet available at ILL.

To verify that our sample is in the low-temperature phase~III, we first measured the $(0\,0\,0.8)$ magnetic Bragg reflection at the base temperature, $T=0.07$\,K, and investigated its magnetic-field and temperature dependencies, which we present in Fig.\,\ref{Fig:AFM_110}. According to the longitudinal $(0\,0\,L)$ scans in panel (a), the peak is centered at $L\approx1/5$, which is very close to the earlier result by Lorenzer~\textit{et~al.} \cite{LorenzerPhD12, DeenUnpublished}, who first observed this peak at an incommensurate wave vector with $L=0.792$. With the application of a magnetic field $\mathbf{B}\parallel[1\overline{1}0]$, the magnetic Bragg intensity is suppressed as shown in Fig.\,\ref{Fig:AFM_110}\,(b). The full suppression is observed at $B_{\rm III}=0.7$\,T, which coincides with the transition to phase~II. Before that, a domain-selection transition occurs at $B_{\rm ds}=0.5$\,T, evidenced by a hysteresis of magnetic intensity measured in increasing and decreasing field. We note that the magnetic domains with the propagation vector $\mathbf{q}_\text{III}\parallel(001)\perp\mathbf{B}$ are favored at the expense of two other domains with $\mathbf{q}_\text{III}\parallel(100)$ and $\mathbf{q}_\text{III}\parallel(010)$ that both form a 45$^\circ$ angle to the field direction, which results in a nearly twofold increase of the $(0~0~0.8)$ Bragg intensity after field cycling. This situation is qualitatively different to the domain selection in the antiferromagnetic (AFM) phase of CeB$_6$, where the application of magnetic field along $[1\overline{1}0]$ suppresses Bragg intensity in the horizontal scattering plane and favors out-of-plane magnetic domains~\cite{EffantinBurlet82, CameronFriemel16}. As a function of temperature, the magnetic Bragg intensity, plotted in Fig.\,\ref{Fig:AFM_110}\,(c), follows an order-parameter-like behavior with an onset temperature $T_{\rm N}\approx0.23$\,K, in good agreement with transport and thermodynamic measurements \cite{ProkofievCusters09}.

\begin{figure*}[t!]\vspace{-5pt}
\includegraphics[width=\textwidth]{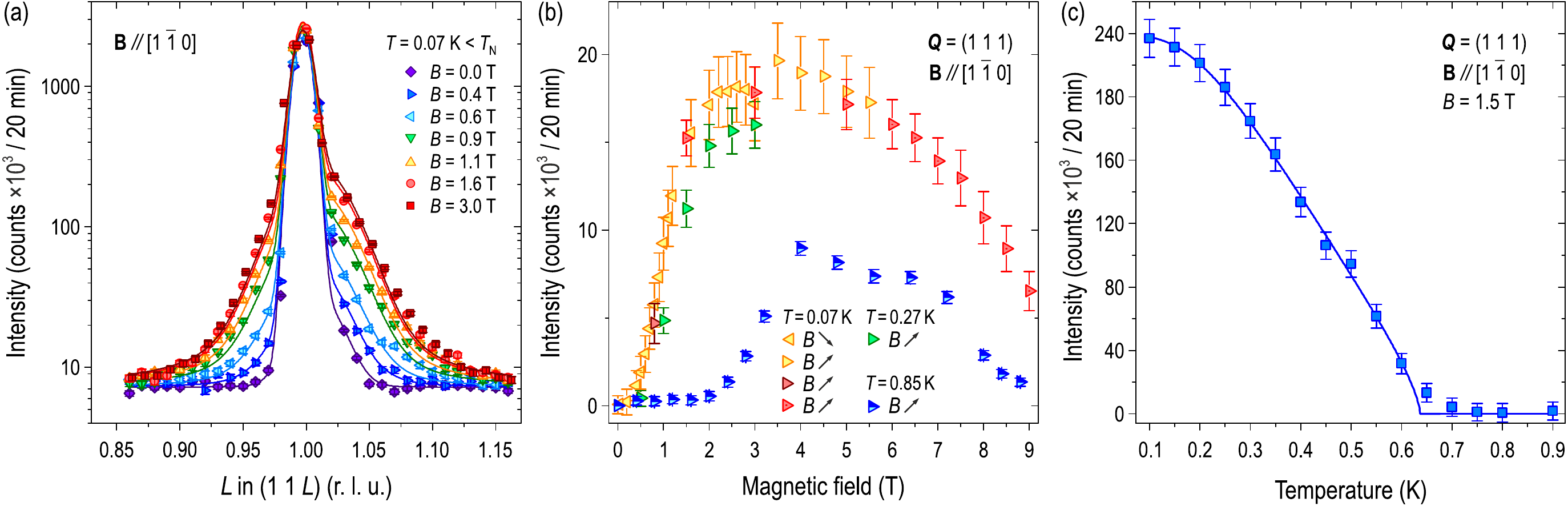}\vspace{-3pt}
\caption{Magnetic-field and temperature dependence of the diffuse magnetic signal that we associate with the order parameter of phase~II, surrounding the $(111)$ structural Bragg reflection. (a)~Typical unprocessed elastic neutron scattering scans measured at $T=0.07$\,K for various fields applied along $[1\overline{1}0]$, fitted with a sum of a Gaussian and Voigt profiles to account for the structural and magnetic contributions on top of a constant background (note the logarithmic intensity scale). (b)~Field dependence of the background-subtracted diffuse magnetic intensity for increasing ($\triangleright$) and decreasing ($\triangleleft$) field, $\mathbf{B}\parallel[1\overline{1}0]$, at three different temperatures. (c)~Temperature dependence of the same intensity, measured in a constant field of 1.5~T. All data points in panels (b) and (c) were obtained from fits similar to those shown in panel (a).\vspace{-2pt}}
\label{Fig:AFQ_110}
\end{figure*}

Now we turn to the discussion of our main result: the observation of an elastic magnetic signal in the vicinity of the $(111)$ wave vector. Because of the peculiar crystal structure of the $R_{3}\mathrm{Pd}_{20} X_{6}$ compounds with interpenetrating simple-cubic and fcc sublattices, this wave vector simultaneously corresponds to the zone corner ($R$ point) for the simple-cubic sublattice and to the zone center ($\Gamma$ point) for the fcc sublattice. As we noted earlier \cite{PortnichenkoCameron15}, the propagation vector of a G-type antiferroquadrupolar (AFQ) order analogous to that of CeB$_6$, residing on the 8$c$~Ce sublattice, would thus overlap with the $(111)$ structural reflection. The corresponding weak magnetic Bragg peak \cite{FriemelLi12} would be thus exceedingly difficult to detect. In Fig.\,\ref{Fig:AFQ_110}\,(a) we present the magnetic-field dependence of the elastic-scattering intensity along the $(11L)$ direction, plotted on the logarithmic intensity scale. In zero magnetic field, only the sharp $(111)$ structural Bragg reflection is observed, whereas magnetic field induces an additional diffuse contribution seen as a much broader peak of nearly field-independent width. This magnetic peak reaches its maximal intensity around 3\,--\,4~T and then starts to decrease again, as shown in Fig.\,\ref{Fig:AFQ_110}\,(b). The signal persists both below and above $T_{\rm N}$, yet in a narrower field range towards higher temperatures. A comparison with the phase diagram in Fig.\,\ref{Fig:PhaseDiagram} clearly establishes that it corresponds to the stability range of phase~II, which also gets narrower upon warming. The temperature dependence of the diffuse intensity, measured in the field of 1.5~T, is shown in Fig.\,\ref{Fig:AFQ_110}\,(c). It shows an order-parameter-like suppression, evidencing a phase transition at $T_{\rm Q}(1.5\,\text{T})\approx0.65$\,K, as expected for phase~II in this field.

We note that the energy analysis allows us to separate the diffuse elastic contribution from the much broader quasielastic fluctuations reported earlier in Ref.\,\citenum{PortnichenkoCameron15}, which were centered at the same wave vector. This ensures that the observed signal corresponds to truly static magnetic correlations that represent the order parameter of phase~II. Its broad width in momentum (as compared to the structural Bragg reflection) indicates that this is a short-range order with a correlation length of about 120\,\AA\ or $\sim$10 lattice constants. The elastic intensity that is absent in zero field and then starts increasing with field after entering phase~II is analogous to the behavior of the AFQ Bragg intensity in CeB$_6$ \cite{RossatMignod87, CameronFriemel16}. However, because of the much lower value of the critical field, $B_{\rm II}$, in Ce$_{3}$Pd$_{20}$Si$_{6}$ the signal saturates and exhibits a maximum at moderate field values ($\sim$\,3\,--\,5~T in Fig.\,\ref{Fig:AFQ_110}\,(b) for $\mathbf{B}\parallel[110]$), which in the case of CeB$_6$ would be shifted to much higher fields. This distinctive field dependence results from the field-induced dipolar moments modulated by the underlying orbital order \cite{EffantinRossatMignod85, TakigawaYasuoka83, CustersLorenzer12}\,---\,a renowned signature of the AFQ state. Hence, our results demonstrate that the order parameter of phase~II in Ce$_3$Pd$_{20}$Si$_6$ represents a short-range version of the AFQ ordering with the same propagation vector and possibly a similar structure as in CeB$_6$, residing on the simple-cubic 8$c$~Ce sublattice.\enlargethispage{3pt}

\begin{figure}[b!]
\mbox{\hspace{-0.01\columnwidth}\includegraphics[width=1.01\columnwidth]{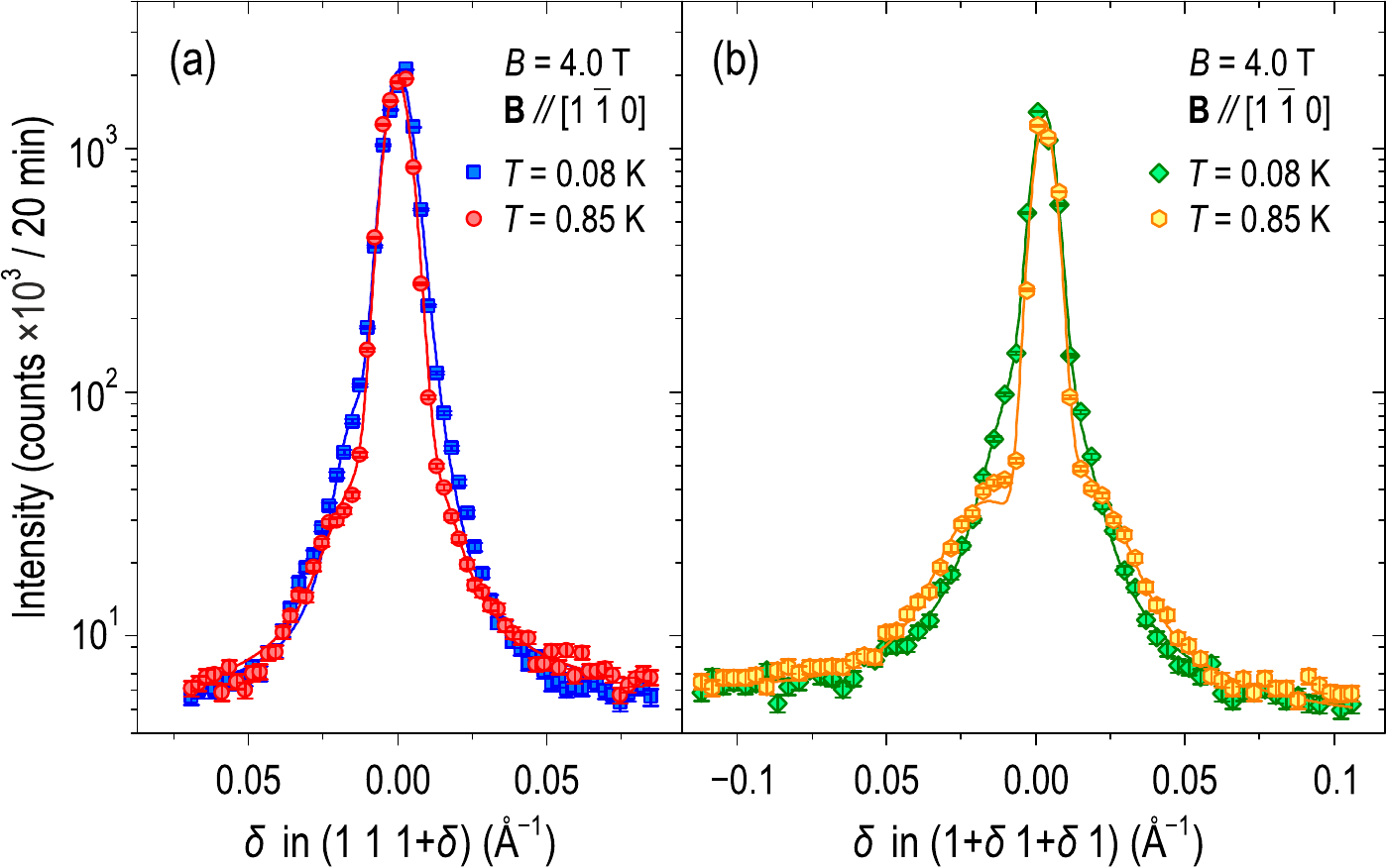}}\vspace{-3pt}
\caption{Variation of the diffuse peak shape with temperature: (a)~along the \mbox{$(1~1~1\!+\!\delta)$} direction; (b)~along the \mbox{$(1\!+\!\delta~1\!+\!\delta~1)$} direction. The magnetic field of 4\,T was applied vertically along $[1\overline{1}0]$, perpendicular to both scan directions.\vspace{-4pt}}
\label{Fig:Fig:AFQ_Tdep}
\end{figure}

A closer inspection of the diffuse peak shape reveals its variation with temperature, which is illustrated in Fig.\,\ref{Fig:Fig:AFQ_Tdep}. The measurements are done in a constant field of 4.0\,T, applied along the same $[1\overline{1}0]$ vertical direction perpendicular to the scattering plane. Along the \mbox{$(1~1~1\!+\!\delta)$} scan direction [panel~(a)], the magnetic intensity is suppressed near the center of the peak upon raising the temperature to 0.85\,K, immediately before the suppression of phase~II. The effect is even more pronounced along the \mbox{$(1\!+\!\delta~1\!+\!\delta~1)$} scan direction [panel~(b)], where we observe a considerable broadening of the peak manifested in the transfer of the magnetic spectral weight away from the central reflection (note the enhanced intensity in the ``tails'' of the peak). These data suggest a tendency towards the formation of an incommensurate modulated quadrupolar structure, as we demonstrate in the following with a dedicated experiment using an orthogonal field orientation.

\begin{figure*}[t!]\vspace{-6pt}
\includegraphics[width=\textwidth]{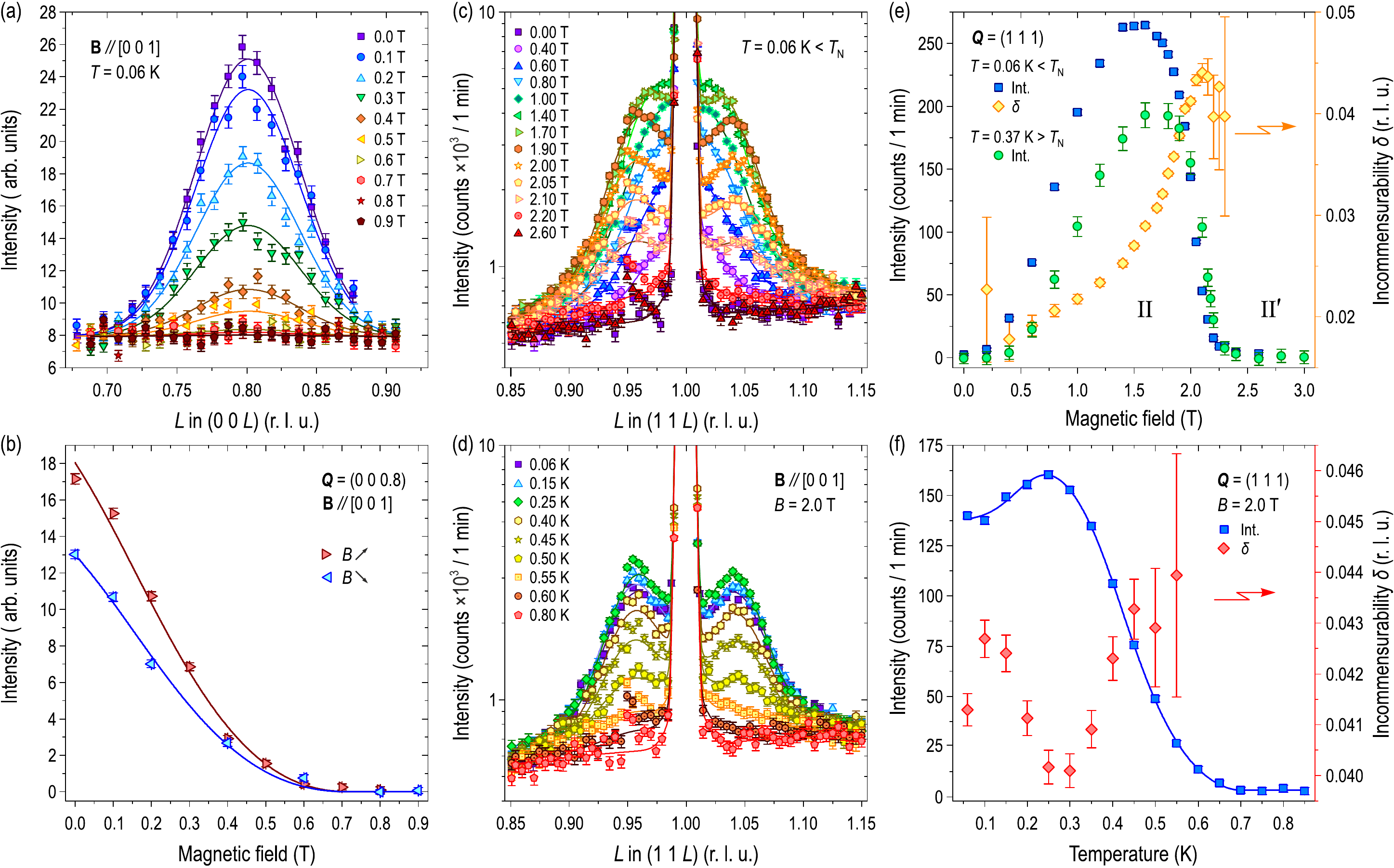}\vspace{-3pt}
\caption{(a)~Magnetic-field dependence of the $(0~0~0.8)$ magnetic Bragg peak for $\mathbf{B}\parallel[001]$, measured at $T=0.06$\,K. (b)~The corresponding field dependence of the Bragg peak amplitude for increasing~($\triangleright$) and decreasing~($\triangleleft$) field, extracted from the Gaussian fits. (c)~Unprocessed elastic-scattering scans along $(11L)$, measured at $T=0.06$\,K for various fields applied along $[001]$. (d)~Temperature dependence of the elastic magnetic intensity at a constant field of 2.0\,T, where the incommensurability of the signal is maximized. (e)~Magnetic-field dependence of the diffuse magnetic intensity at $T=0.06\,\text{K}<T_{\rm N}$ and $0.37\,\text{K}>T_{\rm N}$ (squares and circles, left axis) and of the incommensurability parameter $\delta$ (diamond symbols, right axis). (f)~Temperature dependence of the diffuse magnetic intensity (squares, left axis) and of the incommensurability parameter $\delta$ (diamond symbols, right axis) at a constant magnetic field of 2.0\,T.}
\label{Fig:AFM_AFQ_001}
\end{figure*}

Next, we present similar measurements performed in a horizontal magnetic field, $\mathbf{B}\parallel[001]$. This field orientation corresponds to the lowest critical field for phase~II, $B_{\rm II}^{[001]}\approx2$\,T, and in addition stabilizes the enigmatic phase~II$^\prime$ that is not found for any other high-symmetry field direction. Hence, a natural question that served as an initial motivation for these measurements is whether the diffuse magnetic signal near $(111)$ persists into phase~II$^\prime$ or gets suppressed already at $B_{\rm II}^{[001]}$. Again, we start our presentation with the analysis of the magnetic Bragg peak at $\mathbf{q}_{\rm III}=(0~0~0.8)$, which is shown in Fig.\,\ref{Fig:AFM_AFQ_001}\,(a,b). In contrast to phase~II, the low-temperature phase~III is much more isotropic and nearly insensitive to the direction of the magnetic field. This can be seen from Fig.\,\ref{Fig:AFM_AFQ_001}\,(b), where the full suppression of magnetic intensity occurs at the same field of 0.7~T as for $\mathbf{B}\parallel[110]$. However, in contrast to Fig.\,\ref{Fig:AFM_110}\,(c), the peak intensity in this configuration decreases after cycling the field, which can be explained by a partial suppression of the unfavorable magnetic domain with $\mathbf{q}_\text{III}\parallel(001)\parallel\mathbf{B}$ as a result of the domain selection.

Figures \ref{Fig:AFM_AFQ_001}\,(c) and (d) show the development of the diffuse magnetic signal near $(111)$ in the magnetic field $\mathbf{B}\parallel[001]$ at the base temperature of 0.06\,K, and its temperature dependence in constant magnetic field of 2.0\,T, respectively. Here we observe an increasingly incommensurate response with two broad magnetic satellites centered at $(1~1~1\!\pm\!\delta)$, which surround the structural $(111)$ reflection and move further away from the commensurate position as the magnetic field is increased. From the fitting results shown in Fig.\,\ref{Fig:AFM_AFQ_001}\,(e), it is evident that the magnetic intensity starts to increase upon entering phase~II, reaches a saturation around 1.5\,T, and then rapidly drops to zero across the transition to phase~II$^\prime$, proving that the order parameter of II$^\prime$ is qualitatively different from that of phase~II and is not represented by the observed diffuse signal. In Fig.\,\ref{Fig:AFM_AFQ_001}\,(e), we also present a similar field dependence of the intensity measured at an elevated temperature of 0.37\,K~$>$~$T_{\rm N}$, with a qualitatively similar behavior.

The field dependence of the incommensurability parameter $\delta$, resulting from the fits of the diffuse magnetic intensity, is also shown in Fig.\,\ref{Fig:AFM_AFQ_001}\,(e) with diamond symbols. It demonstrates a clear monotonic increase and gets maximized near the transition between phases II and II$^\prime$, beyond which it can no longer be determined because of the vanishing peak intensity. We have extracted the incommensurability parameter by imposing a constraint on the correlation length at low magnetic fields, i.e. the width of the peak in momentum space at low magnetic fields was extrapolated from its values at higher fields. This assumption results in a finite incommensurability even in the low-field limit, which is however smaller than the peak width and therefore cannot be clearly resolved in the raw data. The low-field datasets in Fig.\,\ref{Fig:AFM_AFQ_001}\,(c) can be equally well described \mbox{with a broader commensurate peak}.

Further, we followed the temperature dependence of the incommensurate magnetic response at the field of 2.0\,T, where the incommensurability parameter is maximized. The corresponding data are shown in Fig.\,\ref{Fig:AFM_AFQ_001}\,(d), and the resulting fitting parameters are plotted in Fig.\,\ref{Fig:AFM_AFQ_001}\,(f) vs. temperature. We observe a non-monotonic behavior of the peak intensity with a local maximum around 0.25\,K, which is consistent with the upturn in the transition field $B_{\text{II\,--\,II}^\prime}(T)$ seen in the phase diagram (Fig.\,\ref{Fig:PhaseDiagram}, dotted line). The incommensurability parameter remains nearly constant as a function of temperature with only minor variations of the order of 10\%, as shown in Fig.\,\ref{Fig:AFM_AFQ_001}\,(f) with diamond symbols.

\vspace{-5pt}\section{III. Summary and Discussion}\vspace{-5pt}

Our results demonstrate the existence of static short-range AFQ correlations propagating along $(111)$, which represent the order parameter of phase~II in Ce$_{3}$Pd$_{20}$Si$_{6}$, seen here for the first time directly in a scattering experiment. With the application of magnetic field, these correlations become increasingly incommensurate and finally vanish across the transitions to either phase~II$^\prime$ or phase~I (for $\mathbf{B}\parallel[001]$ and $\mathbf{B}\parallel[1\overline{1}0]$, respectively). Under the assumption that the field-induced dipolar magnetic correlations are modulated by the underlying orbital order \cite{CustersLorenzer12}, this implies the existence of a rather unusual incommensurate orbitally ordered state whose propagation vector can be continuously tuned by the external magnetic field.

We now discuss possible mechanisms which may lead to this field-tuned incommensurate multipolar order. First we recall that quadrupolar structures with incommensurate modulations have been previously observed, for instance, in PrPb$_3$ \cite{OnimaruSakakibara05, OnimaruKusunose16} and in the so-called ``phase IV'' (IC1) of the solid solution Ce$_{0.7}$Pr$_{0.3}$B$_6$ \cite{KishimotoKondo05, TanakaSera06}. Incommensurate octupole order was also considered as a candidate for the hidden order parameter in URu$_2$Si$_2$ \cite{Hanzawa07}. In the case of Ce$_{1-x}$Pr$_x$B$_6$, it has been suggested that the incommensurability of the $O_{x\!y}$-type quadrupolar order results from the frustration imposed by the competition among the AFQ and AFM exchange interactions in combination with RKKY interactions between the Ce and Pr multipoles, and thermal fluctuations are necessary to stabilize the incommensurate MPO phase. Compared to these cases, the present situation in Ce$_3$Pd$_{20}$Si$_6$ is very unusual in two respects:
\begin{enumerate}\vspace{-5pt}
\item[(i)] The incommensurability varies continuously with field, with no apparent lock-in of the wavevector as opposed to PrPb$_3$ \cite{OnimaruSakakibara05}.\vspace{-5pt}
\item[(ii)] The order is rather short-ranged despite the fact that the compound is stoichiometric, without obvious sources of strong quenched disorder.\vspace{-5pt}
\end{enumerate}
The rather localized nature of Ce orbitals suggests that the incommensurate ordering wave vector has itinerant origin, determined by the Fermi-surface geometry. On the one hand, in the itinerant approach similar structures can be obtained as exotic types of density-wave phases, one prominent example being the incommensurate orbital antiferromagnetism associated with circulating orbital currents \cite{ChandraColeman02} or different kinds of multipolar density waves \cite{RauKee12, IkedaSuzuki12, Das12, ThalmeierTakimoto14}, which were proposed among other scenarios as possible explanations for the hidden-order state in URu$_2$Si$_2$ \cite{MydoshOppeneer11, MydoshOppeneer14}. On the other hand, an alternative scenario, which is more conceivable for our system with strongly localized $f$ orbitals, would involve long-ranged indirect RKKY-type interactions between multipolar moments, which are mediated by the heavy conduction electrons \cite{RossatMignod87, OnimaruSakakibara05, OnimaruKusunose16}. The experimentally established phase diagram also supports this scenario, as it compares remarkably well with earlier theoretical predictions derived from an effective pseudospin model for RKKY-coupled multipoles of the $\Gamma_8$ quartet at the $8c$~Ce site [see Supplementary Information of Ref.\,\citenum{CustersLorenzer12}]. Using a microscopic study and a Ginzburg-Landau analysis, it was shown that, in a finite magnetic field, the AFQ order can induce dipolar AFM order with the same symmetry. Depending on the field direction, this dipolar order is either stable in the entire AFQ phase (generic field direction away from $[001]$) or only in part of it (field along $[001]$). This is in striking agreement with our experimental findings. A stabilization of $T_{\rm{Q}}$ with field, as observed in experiments, is expected for the quadrupolar moment $O^0_2$, which induces a dipolar moment $J_z$ for field along $[001]$. On the other hand, the order in phase II$^\prime$, which remains elusive in the present study, could be of $O_{x\!y}$ type as theoretically suggested, for which no induced dipolar moment is expected \cite{ShiinaShiba97}. This hypothesis remains to be clarified by future experiments.

While incommensurability was not considered initially in the framework of this theoretical model, it is possible that the momentum-space structure of the RKKY interaction, as expressed by the Lindhard function, displays a rather weak momentum dependence near its maximum: Such a situation, arising from a complex underlying band structure, would reflect itinerant frustration. A weak momentum dependence over a range of momenta implies that the position of the maximum can acquire sizeable shifts as function of an applied Zeeman field. Hence, we propose that the RKKY interaction displays a shallow peak at the ordering wave vector with small incommensurability $\delta$ in weak fields, and this peak is continuously shifted to larger $\delta$ with the application of an external magnetic field.

Given the discrete character of the orbital degrees of freedom, a plausible picture for a multipolar state with small incommensurability $\delta$ is that of antiphase domain walls of density $\propto 1/\delta$ in a commensurate background. A periodic arrangement of domain walls yields a sharp Bragg peak. However, these domain walls are naturally susceptible to pinning by defects, which would destroy long-range order and result in a state with short-ranged correlations. The susceptibility to quenched disorder is greatly enhanced by the postulated weak momentum dependence of the RKKY interaction, as this also implies a weak selection of an ordering wavevector. We propose this scenario as a possible explanation of the observed small correlation length.

In summary, we provided direct evidence for field-induced dipolar magnetic correlations in Ce$_3$Pd$_{20}$Si$_6$, experimentally confirming the previously suggested AFQ order parameter of the hidden-order phase II. We suggest that itinerant frustration, reflected in a particularly weak momentum dependence of the RKKY interaction near its maximum wave vector, is responsible for the experimental findings and can explain both the field-dependent incommensurability and the short-range nature of the multipolar order. To verify the scenario of itinerant frustration, detailed band-structure calculations for Ce$_3$Pd$_{20}$Si$_6$ would be required; those are not available to date. Alternatively, photoemission tomography might be used to experimentally determine the low-energy bands which can be used to parameterize the band structure and calculate the Lindhard function, as recently done for CeB$_6$~\cite{KoitzschHeming16}.\vspace{-5pt}

\vspace{-5pt}\section*{Acknowledgments}\vspace{-5pt}

We thank Philippe Boutrouille (LLB) and Eric Bourgeat-Lami (ILL) for technical support during the experiments. This project was funded by the German Research Foundation (DFG) under grant No.~IN\,\mbox{209/3-1}, the Research Training Group GRK\,1621, and the Collaborative Research Center SFB\,1143 (projects C03 and A04). The work in Vienna was supported by the European Research Council (Advanced Grant QuantumPuzzle, No.~227378) and the Austrian Science Fund (Project P29296-N27).

\bibliographystyle{my-apsrev}\bibliography{Ce3Pd20Si6}

\end{document}